\documentclass[]{spie}  
\usepackage{amsmath,amsfonts,amssymb}
\usepackage{graphicx}
\usepackage[colorlinks=true, allcolors=blue]{hyperref}
\usepackage{makecell}
\usepackage{float} 

\title{Jupyter simulations Using NGS Intensity with PyWFS for REDWOODS}

\author[a]{Joel Hurtado}
\author[a]{Benjamin Gerard}
\author[a]{Cesar Laguna}
\author[a]{Dominic Francisco Sanchez}
\affil[a]{Lawrence Livermore National Lab}
\authorinfo{Further author information: (Send correspondence to Benjamin L. Gerard)\\Benjamin L. Gerard.: E-mail: gerard3@llnl.gov}

\begin{document} 
\maketitle

\begin{abstract}
REDWOODS on ShaneAO at Lick Observatory implements a second-stage, 3-sided reflective pyramid wavefront sensor (PWFS), which, under low-light conditions, offers an improved signal-to-noise ratio for deformable mirror commands to correct dynamic atmospheric aberrations. We modeled the REDWOODS limiting natural guide star magnitude. We also investigated performance tradeoffs between two PWFS modes that are available to REDWOODS, one with higher bandwidth error and another with higher aliasing error. We also implemented an experimental setup to image one of the REDWOODS PWFS masks.
\end{abstract}

\section{INTRODUCTION}
Exoplanet imaging is difficult due to light's susceptibility to distortion. While large space-based telescopes are ideal for these applications, they are expensive. Terrestrial telescope performance is impeded by turbulence in the atmosphere, distorting the planar wavefronts as they propagate down to Earth. Wavefront distortions, however, can be corrected with a mirror that can deform such that the reflected light is restored un-aberrated. In order to control the deformable mirror (DM), we need to know where and to what amplitude the distortions are. Wavefront sensors are used to extract information about aberration with a reference light source (natural guide star) that is propagated through a similar path as our object of interest (often the same path for exoplanet imaging). 

REDWOODS\cite{gerard} at Lick Observatory utilizes a pyramid wavefront sensor (PWFS). The design includes a three-sided reflective pyramid mask, advantages of which include achromatic low-light sensitivity\cite{olivier}. In this proceeding we investigate performance of the REDWOODS PWFS.

\section{LIMITING NGS Magnitude}
\subsection{Simulation overview}
\begin{figure}[H]
        \begin{center}
           \includegraphics[scale=0.5]{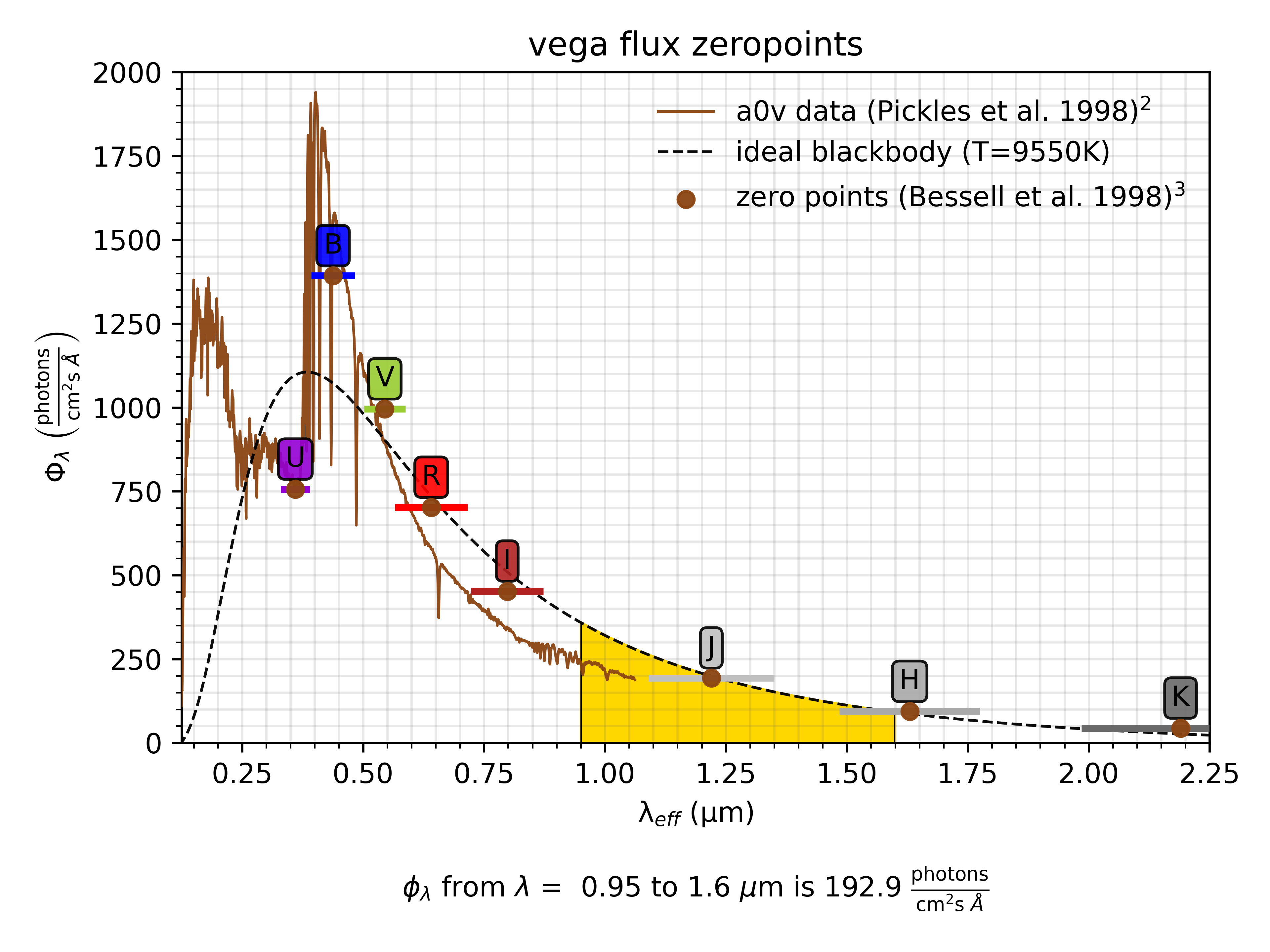}
       \end{center}
           \caption{Fit to the zero flux point for our REDWOODS 3RPWFS spectral bandpass, compared to Vega pickles spectrum and ideal Blackbody.}
       \label{fig:blackbody}
    \end{figure}

To simulate wavefront error propagation, coherent light is used in the middle of our spectral band-pass. Our spectral bandwidth of interest was in the near infrared from $\lambda = 0.95\mu$m to $\lambda = 1.60\mu$m, set by the ShaneAO Shack Hartmann WFS dichroic and the REDWOODS camera quantum efficiency, respectively. The star Vega was used as a model natural guide star from Pickles (1998)\cite{Pickles1998} as shown  in Figure \ref{fig:blackbody} using the equation,
    \[
        Q_d(\lambda) = \frac{2\pi c}{\lambda^4}\left(\frac{1}{e^{hc/\lambda kT}-1}\right)
    \]
This fitted data spectra agreed with Vega flux zero points over standard Johnson-Cousins-Glass band passes\cite{martini} shown in colored boxes in Figure \ref{fig:blackbody}.
    
With the assumptions listed in Table \ref{tab:assumptions}, we calculated the total number of photons over a 1 ms and 10 ms WFS integration time.
\begin{table}[H]
    \caption{Assumptions.}
    \centering
    \begin{tabular}{|c|c|}\hline
        WFS wavelength & 1.275 $\mu$m\\ \hline
        telescope diameter & 3 m\\ \hline
        camera read noise & 30 e$^-$ RON\\ \hline
        quantum efficiency & 0.70\\ \hline
        \makecell{atmospheric\\transmittance} & 30\%\cite{ramirez}\\ \hline
        instrument throughput & 20\% \\ \hline  
    \end{tabular}
    \label{tab:assumptions}
\end{table}
A -2 power law wavefront error phase was used to simulate residual atmospheric aberrations, representing a the REDWOODS first-stage SHWFS correction. This was done by combining a mesh grid of random white noise (phase) and an amplitude parameter to get a complex-valued wavefront. The real component of the Fourier transform was taken of the wavefront to get a pupil plane grid and normalized to set rms wavefront error ($\frac{2\pi}{\lambda}\cdot 100$ nm).

The real value of this simulated aberration was used as the phase amplitude for the wavefront propagated through an aperture. The result was normalized for total photons and Fourier transformed to an image plane so that a pyramid mask could be added and inverse Fourier transformed back to the pupil plane image.

\begin{figure}[H]
    \begin{center}
       \includegraphics[scale=0.5]{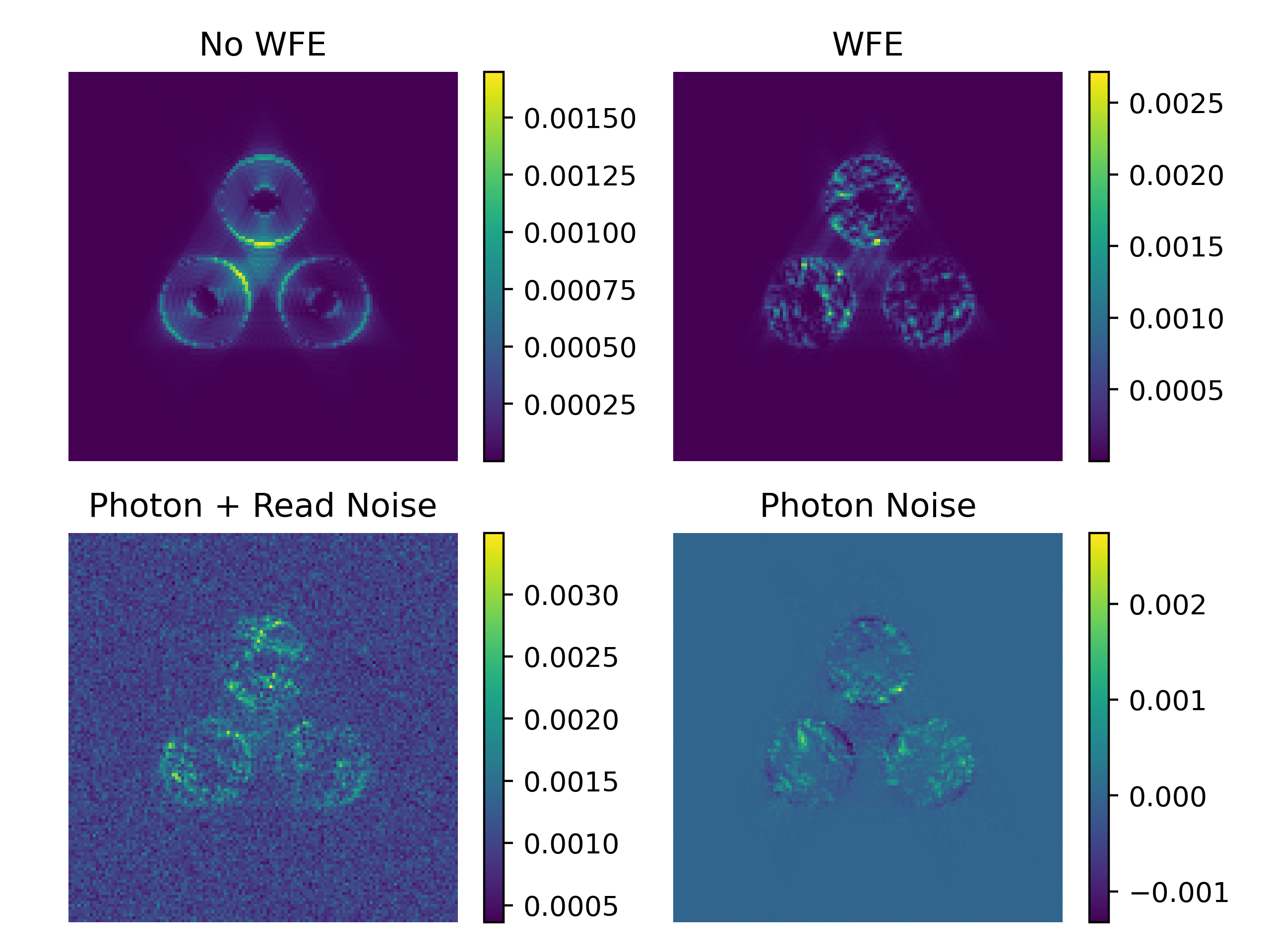}
       \caption[wfs-modes]{\label{fig:wfs-modes} 
    Wavefront simulations with M=7 brightness, 1ms integration time, and WFE amplitude of 0.5 rad rms.}
    \end{center}
\end{figure}

Lastly, read and photon noise were simulated by applying random Poisson distributed values scaled to a read noise level for the entire image or just the sufficiently illuminated pupil plane image for the photon noise case. A resulting example image can be seen in Figure \ref{fig:wfs-modes}.
The case of no wavefront error was simulated to show the reference case for an unaberrated pupil image after applying the pyramid mask. 

\subsection{Signal to noise calculation}
Signal to noise ratio (SNR) is used to determine our ability to correct for aberrations. A low ratio suggests DM commands might erroneously be applied, further amplifying noise. An SNR of 3 roughly corresponds to a 99.998\% confidence in applying the proper DM commands to correct an aberrated wavefront in regards to WFS measurement noise.

A command matrix was computed with a linear reconstructor. Two separate wavefront error phases where used to represent high and low frequency aberrations. These were a Fourier component defined by a $\sin$ function with approximately 10 cycles per pupil and a Zernike quadrafoil (m, n = 4). These wavefronts were applied to a Fraunhofer propagation function as described above. An interaction matrix was constructed by vectorizing the result to a column vector of length equal to the number of pixels in our simulated image. Taking the pseudoinvese of this matrix produced our command matrix (CM). 

Matrix multiplying this CM by a simulated image with WFE produces coefficients of signal strength for a given mode, Fourier and Zernike. Repeating this for an image with the same WFE and additional photon and/or read noise and then taking the difference between the noiseless and noisy coefficients is a measure of the noise strength. The unitless ratio of the noiseless to ``noisy - noiseless'' coefficients is the SNR.

\subsection{Model validation}
\begin{figure}[H]
        \begin{center}
           \includegraphics[scale=0.5]{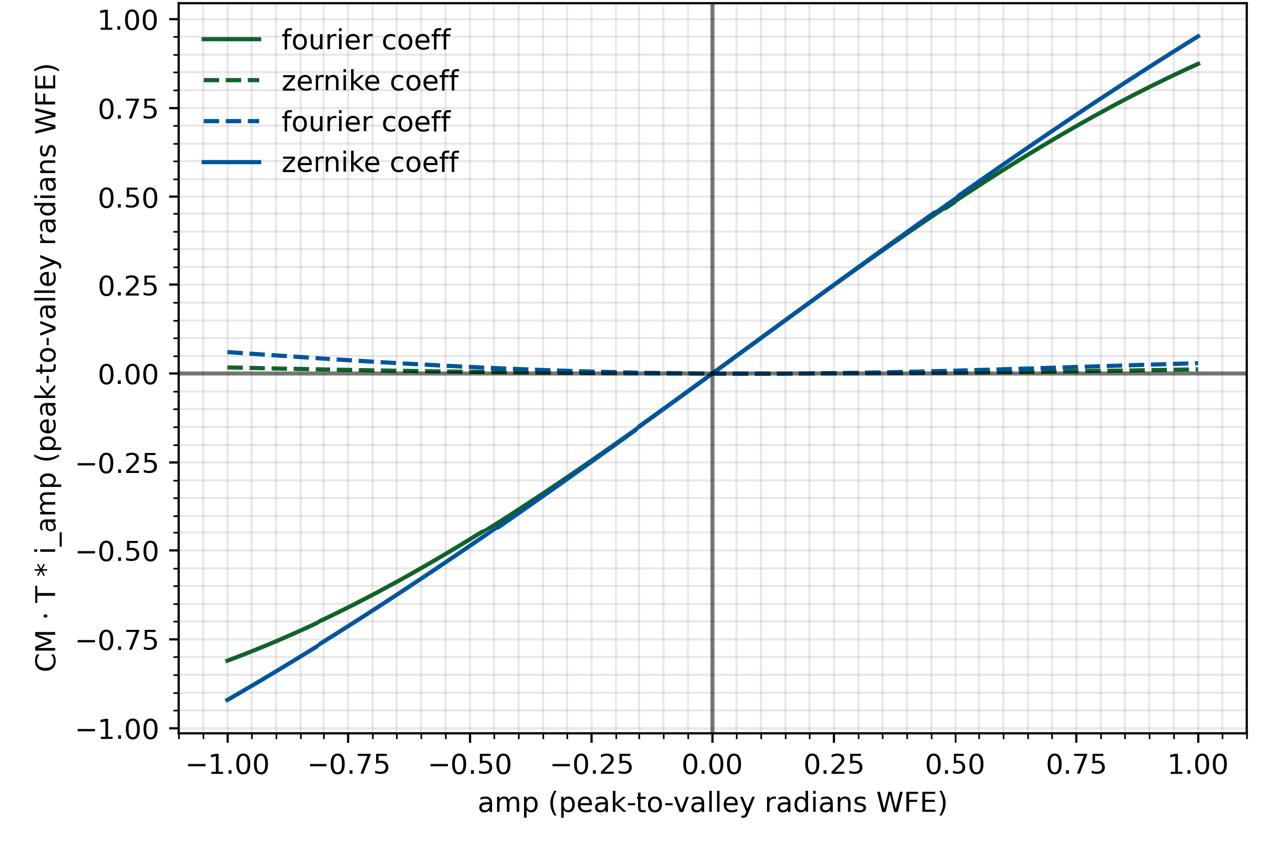}
           \caption[wfs-modes]{\label{fig:linear-test} 
        PWFS linearity simulations for two modes (10 c/p Fourier mode and Zernike quadrafoil, showing the probed mode (solid) and coupling (dashed) when probing the Fourier mode (green) and separately the Zernike mode (blue). }
        \end{center}
    \end{figure}

Two tests were used to gauge the appropriateness of our approach in calculating SNR values. The first was a linear response check. Noiseless images were created with only Fourier and only Zernike wavefront error with varying phase amplitude, from -1.00 to 1.00 peak-to-valley WFE radians in increments of 0.02 radians. Coefficients were generated for each step and plotted as shown in Figure \ref{fig:linear-test}.

\begin{figure}[H]
        \begin{center}
           \includegraphics[scale=0.48]{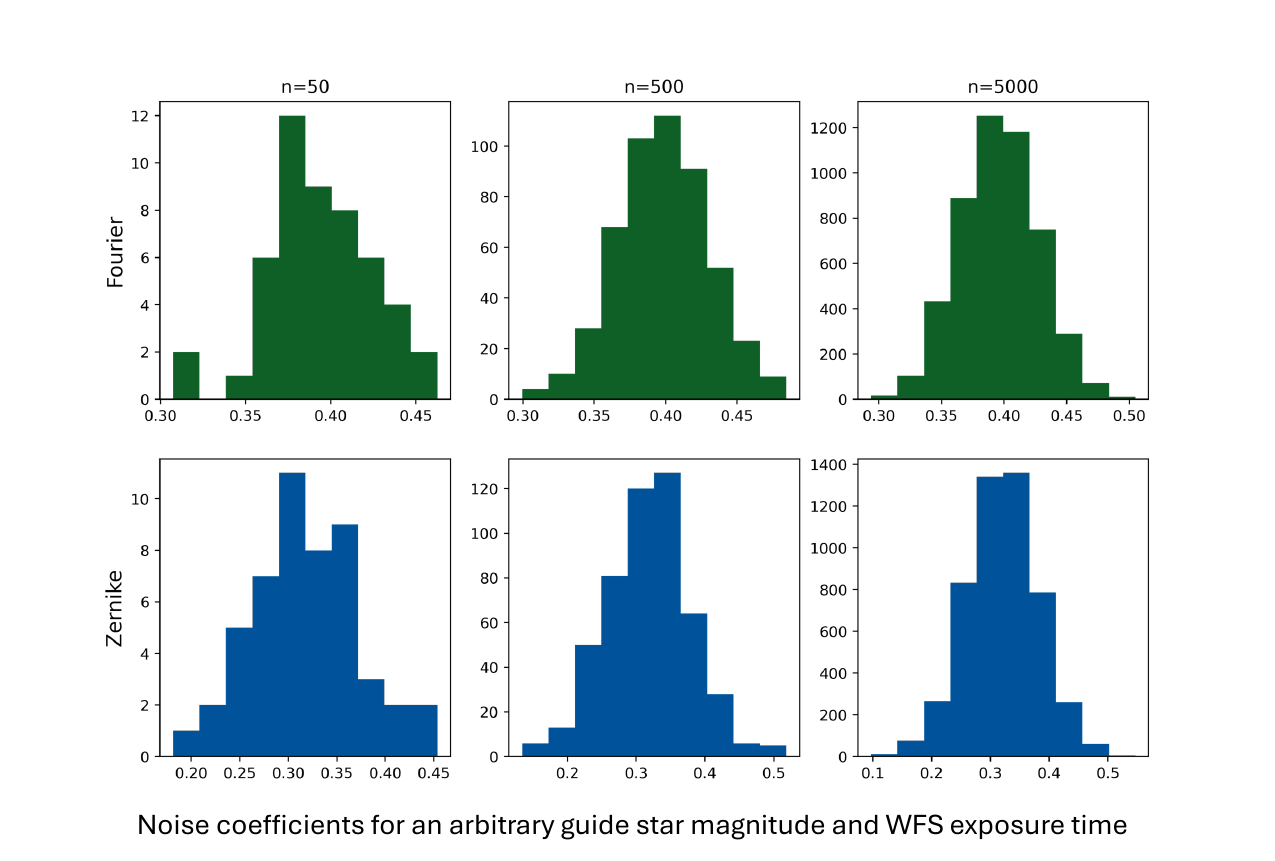}
           \caption[sd-historgrams]{\label{fig:sd-histograms} 
        Normal behavior in noise histogram was observed near n=500 iterations.}
        \end{center}
    \end{figure}

The other test was averaging over a sufficiently large number of iterations with simulated photon and read noise. Since such simulated noise is random, it should display Normal behavior over a sufficiently large enough sample. Iterations of n = 50, 500, and 5000 were taken and the standard deviations of these trials plotted as histograms for noise coefficients for both the Fourier and Zernike coefficients as shown in Fig. \ref{fig:sd-histograms}.
Based on Fig. \ref{fig:sd-histograms}, a sample size of n=5000 was conservatively chosen for calculating SNR values.

\subsection{Results}
\begin{figure}[H]
        \begin{center}
           \includegraphics[scale=0.5]{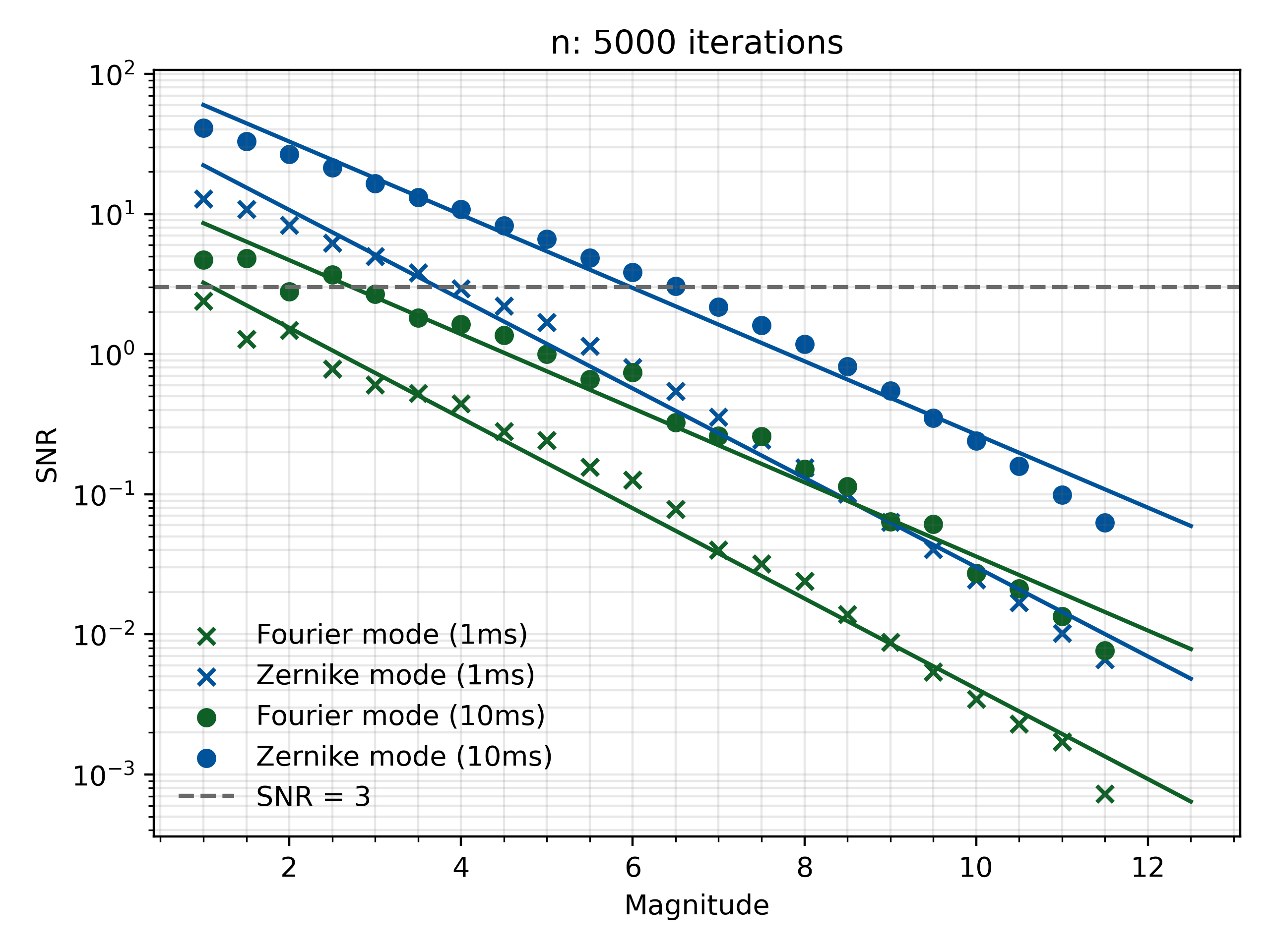}
            \caption[snr-mag-both]{\label{fig:snr-mag-both} 
            SNR calculated in 0.5 magnitude increments for both modes at 1ms and 10ms integration times}
            \end{center} 
    \end{figure}

With a sufficiently large enough sample size (n=5000) and WFE phase amplitude in the linear response domain (generating coefficients no larger than 0.15 radians peak-to-valley), SNR values were calculated for the aforementioned modes with 1ms and 10ms integration times. This was done for NGS magnitudes from 1 to 11.5 in 0.5 increments. A linear best fit line for each scatter plot in log-log space is shown in Figure \ref{fig:snr-mag-both}.\newline

Summarized in Table \ref{tab:summary} are the limiting NGS (SNR$\geq$3) simulated for higher order (Fourier) and lower order (Zernike) aberrations for 1ms and 10ms integration times. This was done for a 3m aperture with 30$^-$e RON camera as implemented in REDWOODS and for a hypothetical case with larger aperture (10m) and improved camera (1$^-$e RON).

    \begin{table}[H]
    \caption{SNR for two modes, two integration times, and two instruments}
    \begin{center}
        \begin{tabular}{| c | c | c | c | c |}\hline
             & Fourier (1ms) & Zernike (1ms) & Fourier (10ms) & Zernike (10ms)\\ \hline
             \makecell{D=3m\\ 30e$^-$ RON \\ (REDWOODS)} & 1.11 & 3.78 & 2.74 & 5.99\\ \hline
             \makecell{D=10m \\ 1e$^-$ RON} & 2.74 & 6.80 & 5.18 & 9.36\\ \hline
        \end{tabular}
    \end{center}
    \label{tab:summary}
    \end{table}
    
\section{BANDWIDTH VS. ALIASING ERROR}
A second REDWOODS pyramid mask is designed with a steeper angle apex such that the PWFS pupil images as seen in Figure \ref{fig:wfs-modes} are shifted outward from the center. This reveals a center pupil image that includes the spatially-filtered super-Nyquist high spatial frequencies of wavefront error, in principle increasing sensitivity of the PWFS by reducing aliasing error but potentially reducing total WFE by increasing bandiwdth error due to increased readout time for a larger image. Below in Figure \ref{fig:shallow-angle} is a simulated image after using a steeper apex angle showing the faint inner-ring.
    \begin{figure}[H]
        \begin{center}
           \includegraphics[scale=0.5]{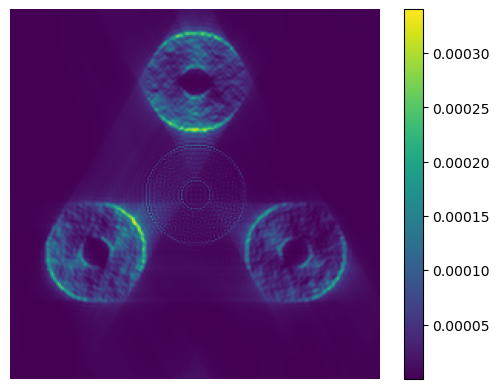}
            \caption[shallow-angle]{\label{fig:shallow-angle} 
            Simulated steeper apex angle REDWOODS PWFS with WFE amplitude of 0.5 rad rms}
        \end{center}
    \end{figure}
\subsection{Bandwidth Error}
REDWOODS uses a CRED2 LITE camera with a sensor size of 640x512 pixels. The frame rate of different regions of interest varies by number of rows and columns. The top left single pixel maximum readout rate is around 32,000 fps whereas the full frame is around 600 fps. The decrease in frame rate is non-linear.

While the steeper apex angle allows for potentially reduced aliasing, it also increases the footprint of the image. Combined with ongoing work that has recently benchmarked REDWOODS upper limit computational latencies of 574 $\mu$s and 589 $\mu$s for the steeper and shallower apex angle PWFSs, respectively (S. Cetre, private communication) and due to this upper limit assuming negnigiglbe camera or DM frame transfer or DM settling time, for the the shallower apex angle PWFS the limiting frame rate is 10394 fps, resulting in a end-to-end system latency of $\tau = 2.10$ ms, while for the steeper apex angle PWFS, the limiting frame rate is 7534 fps, resulting in $\tau=2.13$ ms. 

With these time delays and utilizing open-source code\cite{AO_control_guis}, we calculate the -3dB bandwidths to be 32.0Hz and 32.4Hz for the shallow and narrow PWFS respectively, applying these two different system latencies and in both cases adjusting integrator control gains until a 45$^\circ$ phase margin minimum stability limit is met to determine these above-mentioned -3dB bandwidths. This represents how fast the AO system can correct for atmospheric distortion. 

Assuming average wind velocity of $v_w$ = 10 m/s and Fried parameter $r_0$ = 10 cm at 500 nm, we calculated using equations 2.71 and 2.27 from Principles of Adaptive Optics\cite{Tyson} a Greenwood frequency of 14.0 Hz.
    \[
        f_G = 0.43 \frac{v_w}{r_0}
    \]

Bandwidth error for each case was calculated using equation 3.32\cite{Tyson}

    \[
        \sigma^2 = \left(\frac{f_G}{f_{3 \text{dB}}}\right)^{5/3}
    \]
The different is bandwidth error between both masks was negligible (0.005 rad). This suggest there isn't a significant tradeoff for using the steeper apex angle PWFS mask.

\subsection{Aliasing Error}
Another tradeoff when considering using a shallower or steeper apex angle PWFS is aliasing. Aliasing occurs when high spatial frequency distortions are under-sampled in the AO system causing them to appear as lower order aberrations. To quantify aliasing error, we need to calculate coefficients similarly to Section 2.2, with the two different pyramid masks. Each PWFS needs it's own simulated command matrix (pseudo-inverse of the interaction matrix) since the patterned footprint of the shallow apex angle mask is 1.5 times smaller than the steeper apex angle mask, which relates to the 20 $\mu$m etching depth limitation of the reactive ion etching method used to fabricate the PWFS masks.

Additionally, an algorithmic anti-aliasing Butterworth filter is used in simulation to filter out higher order aberrations for the entire pupil image. Just as in Section 2.3, linearity plots as a function of WFE amplitude were used to check our coefficients were in the correct units and calculated using the appropriate command matrix.

These coefficients were generated (n=500) times and averaged. The results are summarized in the below in Table \ref{tab:aliasing} in units of radian WFE.
    \begin{table}[H]
        \caption{Results of our aliasing simulations, showing average coefficient strength in radians for various simulated cases.}
        \centering
        \begin{tabular}{|c|c|c|c|c|}\hline
            WFE mode & \makecell{shallow apex angle\\no algorithmic filter} & \makecell{shallow apex angle\\antialias filter} & \makecell{steep apex angle\\no algorithmic filter} & \makecell{steep apex angle\\antialias filter} \\\hline
            fourier & 0.4063 & 0.3851 & 0.4786 & 0.4782\\
            zernike & 0.3661 & -0.1930 & 0.3935 & -0.2439\\\hline
        \end{tabular}
        \label{tab:aliasing}
    \end{table}
Table \ref{tab:aliasing} shows that the aliasing error---the difference between using an anti-alias and no algorithmic filter for a given PWFS mask and mode---is more significant than the effects of bandwidth error for the Zernike mode for both PWFS masks and for the Fourier mode for the shallow apex mask. Interestingly, the Fourier mode for the higher apex angle mask is almost negligible (0.0004 radians), demonstrating as expected the reduction of aliasing due to the natural spatial filtering effect of this mask.

\section{LAB TESTING}
We obtained an image after building an optical system at LLNL using the shallow angle REDWOODS PWFS mask, shown in Figure \ref{fig:PWFS}. Subsections in this section describe how this image was obtained.
\subsection{Design criteria}
This 4f optical system was designed such that the final beam diameter took up one-third of the camera (FLIR BFLY-U3-23S6M) sensor. This was chosen since the expected pupil image after the PWFS would have three rings. Given a pixel pitch of 5.86 $\mu$m physical sensor size of 640x512 pixels, we determined D2 should be less than 2.5 mm. 

We design the PWFS mask f number to be f/30 to match REDWOODS. Given these constraints, L2 focal length was chosen to be 200 mm and L3 to be 75 mm. Using the equation below,\cite{field_guide}
    \[
        \text{f-number} = \frac{\text{focal length}}{\text{beam diameter}}
    \]
D1 was about 6.6mm and since $D_2 = D_1 \frac{L_3}{L_2}$, D2 must be 2.475 mm, just under the maximum 2.5 mm.

\begin{figure}[H]
        \begin{center}
           \includegraphics[scale=0.25]{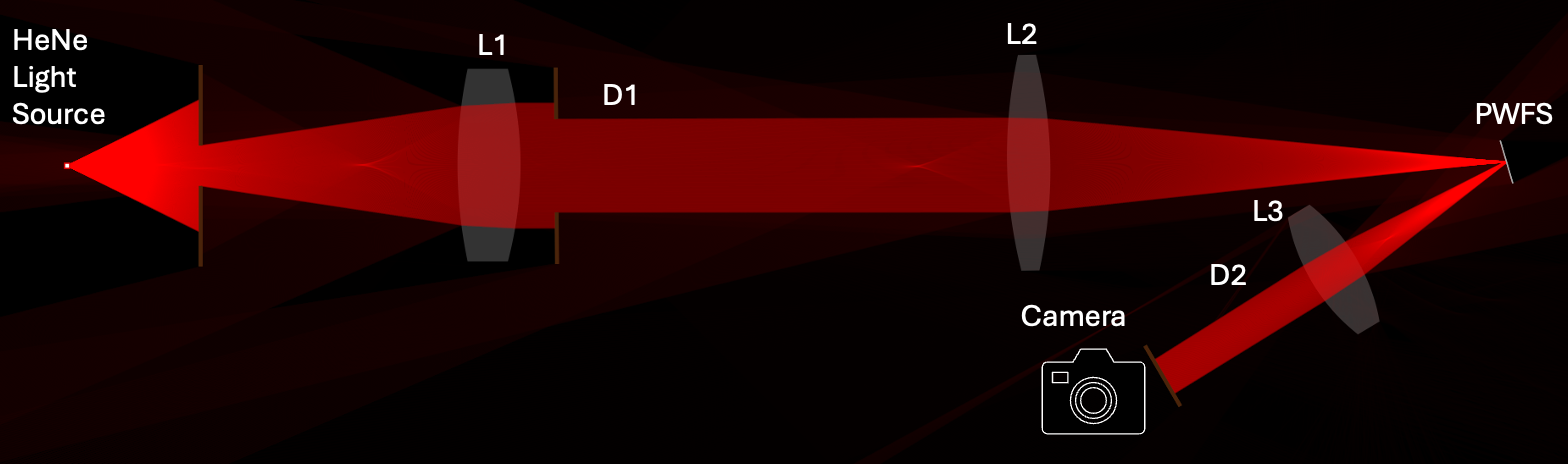}
            \caption[shallow-angle]{\label{fig:setup-diagram} 
            Diagram of experimental setup, not depicted to scale (actually aligned in a 4f configuration). Figure made with Ref. \citenum{RayOptics}.}
        \end{center}
    \end{figure}

Table \ref{tab:components} summarizes the focal length of the $\varnothing$1" convex lenses and beam diameters noted above.
    \begin{table}[H]
    \caption{components}
    \begin{center}
        \begin{tabular}{| c | c|}\hline
            component & focal length (L) or size (D) \\\hline
             L1 & 100 mm\\
             L2 & 200 mm\\
             L3 & 75 mm\\
             D1 & 6.6 mm\\ 
             D2 & 2.48 mm\\\hline
        \end{tabular}
    \end{center}
    \label{tab:components}
    \end{table}
The setup is further illustrated in Fig. \ref{fig:setup-diagram}.
    
\subsection{Alignment}
The HeNe laser light source was mounted by post on the bench top via a fiber optic cable guide. This diffuse light source was collimated with L1. But first to ensure proper alignment, two iris pupil stops adjust to be the same height with a height gauge. The first pupil stop was placed right after the light source. The second was initially placed at the far end of the bench, about 120 cm away. The first stop was closed to get the smallest beam possible. A white card on a magnetic mount was placed right after the stop and marked. It was then placed after the second pupil stop and adjusted so that the beam landed on the same marking. 

After the two stops were aligned the collimating lens L1 was positioned 100 mm away from the light source (one focal length. Alignment was checked similarly as before. We also used the back reflections from the L1 that create a focus spot near the first iris for alignment. Using a white card, we were able to fine-tune the positioning of L1 such that this back reflected focus point landed at the center of the first stop. Collimation was verified through the use of a magnifying shear plate by checking for parallel lines to a reference on the instrument.

The next focusing lens L2 was aligned similarly to L1. The camera was placed at the focal point (where the PWFS is shown) to image the point spread function and verify sharp distinct rings. After that was verified the second aperture stop was moved 100 mm downstream from L1 to set our desired beam diameter.

\begin{figure}[H]
        \begin{center}
           \includegraphics[scale=0.25]{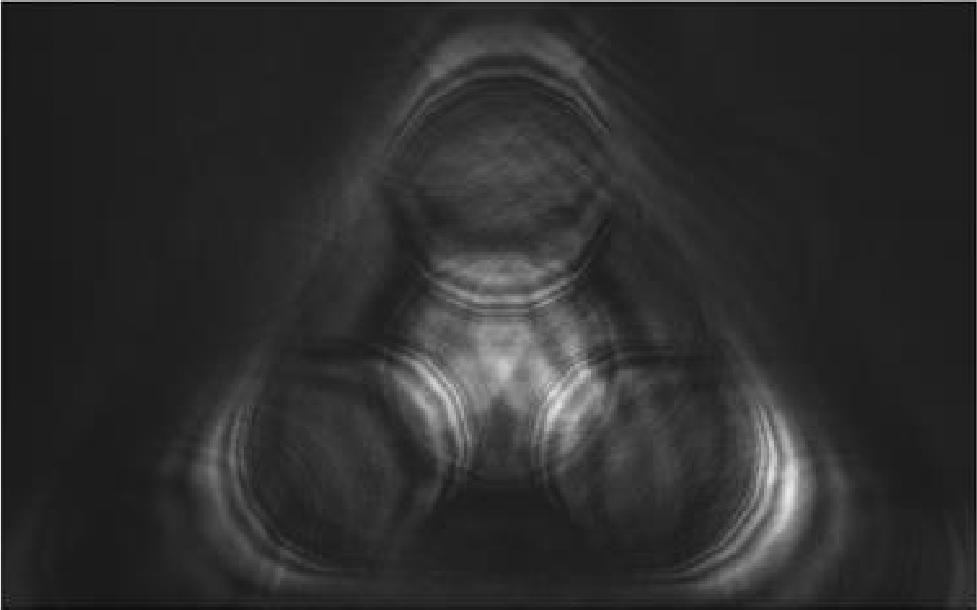}
            \caption[shallow-angle]{\label{fig:PWFS} 
            REDWOODS PWFS (shallow apex angle) image.}
        \end{center}
    \end{figure}

The PWFS was placed at the focal point and aligned to reflect the beam approximately 15° away. The final lens L3 was placed one focal length away from the PWFS and collimation was checked through the use of a shear plate as mentioned above with the patterned PWFS mask translated off of the beam such that the PWFS mask acted as a fold mirror during these measurements. The camera was used to image the pupil (also initially with the PWFS mask as a fold mirror) for camera positioning ane eventually fine-tuning the PWFS z position. The best image was recorded as shown in Figure \ref{fig:PWFS}.

\section{CONCLUSION/FUTURE WORK}
In this analysis we have shown that for the PWFS AO system on REDWOODS, the limiting NGS brightness was around a magnitude 6 with improvement with a larger telescope and less noisy camera. Bandwidth error for a shallow more sensitive PWFS was negligible but aliasing error is more significant. Finally we physically imaged a REDWOODS PWFS mask using a 4f optical setup at LLNL. 

For future work, it would be interesting to investigate why our limiting NGS magnitude is relatively poor (e.g. corrections in our model throughput assumptions). Additionally, REDWOODS will take into account more than just two WFE modes. It then makes sense to incorporate into the model a more representative range of WFE modes to correct for aberrations. When calculating WFE modal coefficients, the entire image was used, including relatively un-illuminated regions that did not contain information about the WFE and, e.g., for the high apex-angle mask including the central pupil that potentially degraded it's optical anti-aliasing effects. An algorithmic mask could thus be implemented to filter out some pixels to further improve the SNR calculation as well as the aliasing error.

The experimentally measured PWFS image exhibited unexpected ringing even after refining optical setup alignment. It would be useful to use manufacturer PWFS metrology data in the simulations to compare with experimental data to help understand the source of ringing in the images.

\acknowledgments

This work was performed under the auspices of the U.S. Department of Energy by Lawrence Livermore National Laboratory under Contract DE-AC52-07NA27344. This document number is LLNL-PROC-2010335.

\bibliography{report} 
\bibliographystyle{spiebib} 

\end{document}